# Using Deep Convolutional Neural Networks to Diagnose COVID-19 From Chest X-Ray Images

Yi Zhong

(School of Information and Communication Engineering, Beijing University of Posts and Telecommunications, Beijing)

**Abstract:** The COVID-19 epidemic has become a major safety and health threat worldwide. Imaging diagnosis is one of the most effective ways to screen COVID-19. This project utilizes several open-source or public datasets to present an open-source dataset of COVID-19 CXRs, named COVID-19-CXR-Dataset, and introduces a deep convolutional neural network model. The model validates on 740 test images and achieves 87.3% accuracy, 89.67 % precision, and 84.46% recall, and correctly classifies 98 out of 100 COVID-19 x-ray images in test set with more than 81% prediction probability under the condition of 95% confidence interval. This project may serve as a reference for other researchers aiming to advance the development of deep learning applications in medical imaging.

**Keywords:** COVID-19 virus; X-Ray images; Deep Convolutional Neural Network; Deep Learning

## 1 Introduction

In late 2019, a sudden new coronavirus (hereafter referred to as COVID-19) swept through Wuhan. Soon the epidemic spread rapidly throughout China. Until June 29, 2020, there were 10 million cases and nearly 500,000 deaths[1] of COVID-19 have now been reported globally. The COVID-19 epidemic has become a major safety and health threat worldwide. One of the key steps to curb the spread of the COVID-19 epidemic is effective screening of suspected infected persons to prevent the virus from spreading further among the population. The most accurate screening method for new coronary pneumonia is nucleic acid detection, that is, reverse transcriptase-polymerase chain reaction (RT-PCR) detection is taken from nasopharyngeal swabs, sputum and other lower respiratory tract secretions, blood, and stool samples[2]. Although RT-PCR testing is the gold standard for COVID-19 screening, RT-PCR testing is a very complicated, time-consuming and labor-intensive process. Not all health institutions in all regions have such testing conditions. At the beginning of the outbreak, the number of RT-PCR detection reagents is very limited. In the background of the global epidemic, some countries and regions still do not have a sufficient number of RT-PCR testing kits.

After the outbreak of the COVID-19 epidemic, the National Health Comission of the People's

Republic of China in the "New Coronavirus Infected Pneumonia Diagnosis and Treatment Program (Trial Fifth Edition)"[3] for the first time included suspected cases with chest imaging features into clinical diagnosis cases, that is The suspected cases of new coronary pneumonia can be diagnosed based on imaging features. The first and main imaging test for the diagnosis of new coronary pneumonia is High Resolution Computerized Tomography(HRCT) of the chest. Considering that the testing capacity and human resources of health care institutions in some regions are limited, and the cost of HRCT testing is relatively high, the radiation is relatively large, especially for patients who are suspected of being infected with new coronary pneumonia but are ultimately undiagnosed(maybe other infection like bacterial pneumonia or other virus infection). This is undoubtedly a great price for patients. Therefore, for these areas where medical resources are insufficient or the medical resources have been saturated during the epidemic, it is of great significance to use Chest X-ray (CXR) for the initial screening of the COVID-19. This project designs and implements a deep convolutional neural network to classify CXR images, which can be used as an auxiliary method to help artificial X-ray imaging rapid screening of the COVID-19.

The paper is organized as follows: Section 2 introduces the COVID-19 imaging features and the related work of deep learning in COVID-19 imaging diagnosis; Section 3 introduces the dataset source of this project and the main research and optimization methods, then proposes a deep convolutional neural network model; Section 4 introduces the experimental environment that the project depends on and the evaluation index results of the model on the training set and testing set; Section 4 summarizes the project and its shortcomings, and expounds the outlook for future work; Finally section are the references of this paper.

## 2 Research on COVID-19 imaging diagnosis

### 2.1 COVID-19 imaging features

According to the description of the imaging diagnosis in the "New Coronavirus Infected Pneumonia Diagnosis and Treatment Program (Trial Seventh Edition)"[2], the chest imaging of patients infected COVID-19 "In the early stage, there are multiple small patch shadows and interstitial changes, with obvious extrapulmonary zone. Then it develops multiple ground glass shadows and infiltration shadows of the lungs. In severe cases, lung consolidation may occur, and pleural effusions are rare." According to the analysis of the recently published literature by Deng et al[4], milled glass shadow and solid change shadow are the most common imaging signs, and lesion It is mainly located peripherally and subpleurally, along the bronchovascular bundle, and is diffusely distributed in both lungs as the disease progresses. A Normal chest X-ray image has been shown in Figure 1, and a chest X-ray image of a COVID-19 patient has been shown in Figure 2, some COVID-19 imaging features can be observed in comparison with Figure 1 and Figure 2 .

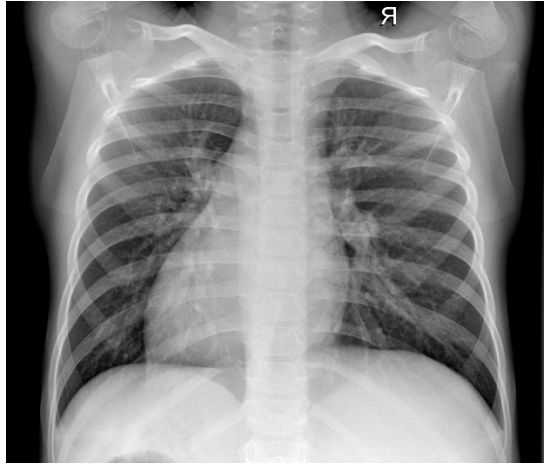

Figure 1  A Normal chest X-ray image

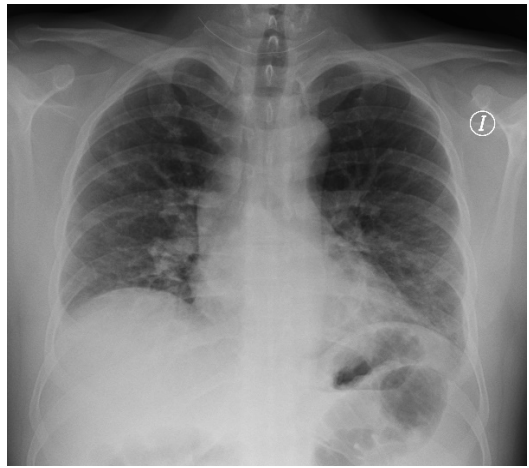

Figure 2  A chest X-ray image of a COVID-19 patient

## 2.2  Related work of deep learning in COVID-19 imaging diagnosis

In recent years, artificial intelligence, especially deep learning technology, with its powerful data analysis and feature recognition capabilities, has become prominent and clinically recognized in the diagnosis and recognition of lung diseases such as chest X-ray quality control, lung nodule segmentation and detection, and lung cancer screening[5]. The study shows that AI has a sensitivity of 71%, specificity of 95%[6] and far better recognition speed than humans in distinguishing normal from abnormal chest radiographs on CXR image classification.

In the middle of the epidemic in China , Alibaba DAMO Academy cooperated with Alibaba Cloud and then developed a set of new AI diagnostic technology for clinical diagnosis of COVID-19, which can accurately interpret CT images of suspected cases of COVID-19 within 20 seconds and the accuracy of analysis results reached 96%[7], and the system first landed in Xiaotangshan Hospital Zhengzhou, China on February 16 for auxiliary clinical diagnosis. The effectiveness and usefulness of AI in medical imaging is evident, providing strong technical support to contain the spread of the epidemic.

The current imaging diagnosis of COVID-19 is mostly based on CT images, due to the high rate of misses on chest X-rays for early lesions or predominantly milled glass density changes. The advantages of CT especially HRCT, are its high spatial resolution, freedom from interference from structures beyond the level, and multi-planar, multi-directional display of lesion detail through post-processing techniques. Xu et al[8]. proposed a 3D deep learning model to segment candidate infection domains from the lung CT image set and further calculate confidence scores using a regional attention mechanism model, capable of accurately screening COVID-19 by lung CT images. The Section 1 in this paper mentions the continued importance of x-ray images for screening for COVID-19, El-Din Hemdan et al[9]. proposed a deep learning classifier framework called COVIDX-NET for diagnosing COVID-19 in CXR, the paper also validated seven different DCNN models such as VGG19 and Densenet201 and found that VGG19 and DenseNet classification performed better. Wang Linda and Wong Alexander[10] presented a deep convolutional neural network named COVID-Net to detect COVID-19 cases from chest X-ray images and analyzed how COVID-Net makes predictions using interpretable methods, in order to gain insight into the key factors associated with COVID-19 cases that can help clinicians perform better screening.

## 3 Methods

### 3.1 COVID-19 CXR Dataset

As of today, the number of publicly available COVID-19 image datasets remains small. Some middle COVID-19 dataset like Joseph Paul Cohen et al[11]. provide in their paper. This paper proposes an open source COVID-19 dataset named COVID-19 CXR Dataset, which can be obtained from this Github link(https://github.com/ZY-ZRY/COVID19-CXR). COVID-19-CXR-Dataset has a total of 6,354 CXR images, including test set and training set, divided into three categories: COVID-19, Normal and Pneumonia. Category COVID-19 are CXR images of patients infected COVID-19; Category Normal are normal CXR images; Category Pneumonia are CXR images of patients infected other kinds of pneumonia. The distrubution of the category in the COVID-19 CXR Dataset is shown in Table 1.

The CXR images in COVID-19-CXR-Dataset are obtained from these open source and public datasets:
- COVID-19 image data collection[11]
- Figure 1 COVID-19 Chest X-ray Dataset Initiative[12]
- ActualMed COVID-19 Chest X-ray Dataset Initiative[13]
- Chest X-Ray Images (Pneumonia)[14]

Special thanks to Joseph Paul Cohen [11-14] and others for their contributions to the COVID-19 and pneumonia datasets!

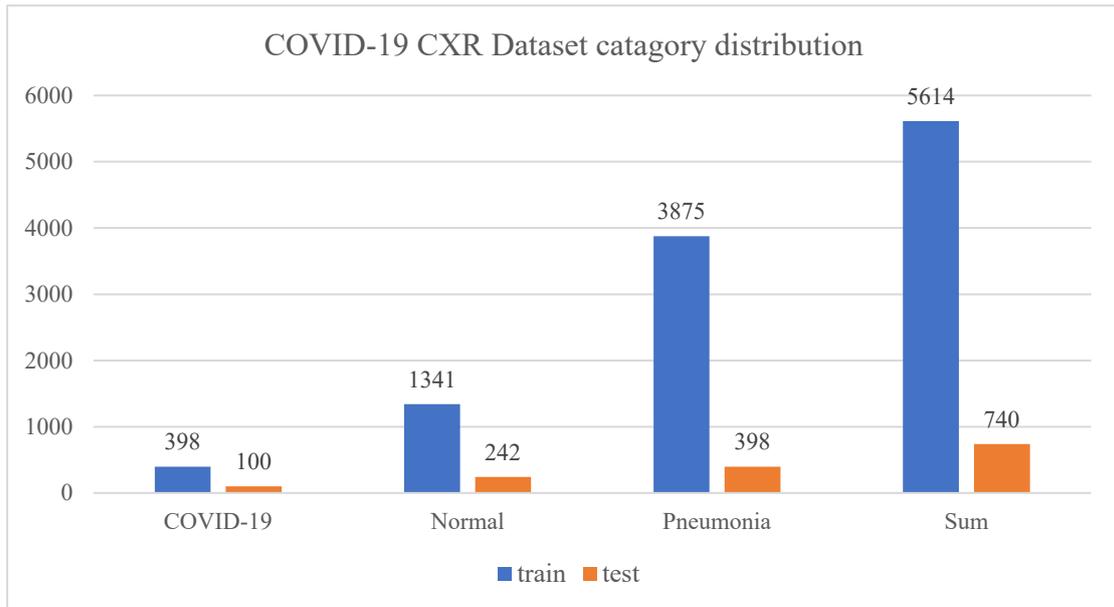

Table 1  COVID-19 CXR Dataset catagory distribution

### 3.2 Data pre-processing, data augmentation and category balance

During to the CXR images in COVID-19 CXR Dataset are obtained from a few of public and open source datasets, the image format, image size, and location of the chest range image capture (anterior/posterior/sideways) in the dataset are inconsistent, some images even have distinct lesion labeling information, which interfered significantly with image recognition. So before training, I have pre-processed the image data firstly, remove the CT images in the dataset and the CXR images with lesions. In addition, the number of CXR images taken on the side is small, and they are also removed. Then I convert the CXR images into jpeg/jpg format. Finally the images are uniformly resized as $80 \times 80$ before they are imported into the model.

There are only 6,354 CXR images in COVID-19 CXR Dataset, the data is not rich enough. From the chart statistics in Table 1, we can see that there are obvious data imbalances in the three categroies of CXR images in the training set. In view of the lack of data, some data enhancement methods were used in the project, for example, random horizontal flip (vertical flip is meaningless for medical images), random brightness adjustment, random contrast adjustment, random saturation adjustment and random cropping, etc. These methods could enrich the dataset, reduce the situation of overfitting the model, and enhance the robustness of the model. In view of the imbalance of data categories, the gradient descent of the loss function of the project during training introduces class weights, which makes the categories with less data have greater losses in training. So the model would pay more attention to the categories with less data.

## 3.3 Deep convolutional neural network model in this paper

Deep convolutional neural network is an intuitive and powerful network architecture in deep learning, and is widely used in pattern recognition and image classification tasks. In recent years, with the further development of deep learning technology, more and more efficient DCNN models have been proposed one after another, like VGG, ResNet, DenseNet, EfficientNet etc. These DCNNs perform well in image classification tasks, making it possible for computers to perform better than humans in visual classification.

VGG[15] networks only use small convolution kernels and increase the number of network layers to improve the accuracy in the classification tasks. There is no doubt that increasing the depth of the network allows the model to better fit complex data with higher dimensions, and the use of small convolution kernels can greatly reduce the number of parameters. Shallow networks usually have rich semantic information, but VGG does not make the most of this semantic information.

This paper proposes a deep convloutional neural network model based on the VGG16. Limited by my poor device(only have 2GB graphic card memory), I have reduced the number of layers and filters to make sure my graphic card memory won't explode(After that I seem to find that increasing the number of network layers and filters does not significantly improve the metrics). In order to make full use of the semantic information of the network, I also made some other changes, like add GlobalAveragePooling layer after Maxpooling and concatenate these independent neurons.

The model This model mainly consists of four sets of convolution-pooling networks, a concatenate layer and two flatten layers. The model has 63,148,835 parameters, and its complete structure is shown in Figure 3.

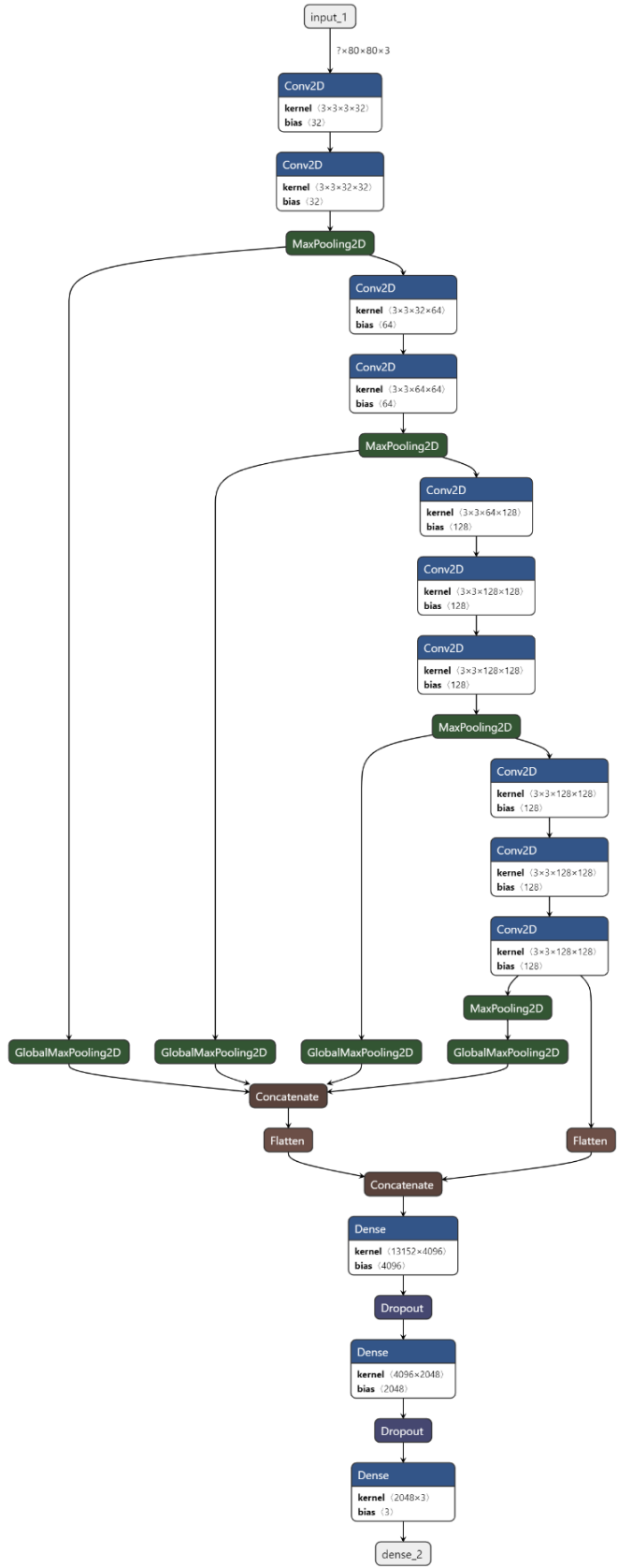

Figure 3 DCNN model proposed in this paper

## 3.4 Loss function and metrics

During the training, I used CategoricalCrossentropy as my loss function, For every input sample $i$ and its category label $j$, has

$$Loss_i = -\sum_j t_{i,j} \log(p_{i,j})$$

where $t_{i,j}$ is the label for sample $i$, indicating that the category of sample $i$ is $j$; $p_{i,j}$ is the probability that sample $i$ is predicted to be category $j$.

In particular, for one-hot encoding labels in single-label classification problems, the label $t_{i,j}$ for each sample $i$ has only one element of one and the rest are zeros, then the loss function becomes

$$Loss_i = -\log(p_{i,j})$$

so the overall loss function is

$$Loss = -\frac{1}{n}\sum_i \log(p_{i,j})$$

where $p_{i,j}$ is the probability that sample $i$ is predicted to be sample $i$'s actual category $j$.

To better evaluate the validity of the model, this project uses a diversity of evaluation metrics. First of all, the concept of a confusion matrix is first introduced here, and the composition of the confusion matrix is shown in Table 2.

| | Confuse Matrix | |
|---|---|---|
| | Predict Positive | Predict Negative |
| Actually Positive | True Positive (TP) | False Negative (FN) |
| Actually Negative | False Positive (FP) | True Negative (TN) |

Table 2  Confuse Matrix

The confusion matrix consists of the following four indicators:

- True Postive (TP): the number of correctly classified samples as positive category
- False Negative (FN): the number of wrongly classified samples as negative category but actually belong to postive categroy
- False Positive (FP): the number of wrongly classified samples as positive categroy but actually belong to negative category
- True Negative (TN): the number of correctly classified samples as negative category

With the four indicators of the confusion matrix, three evaluation indicators Accuracy, Precision, and Recall (also called Sensitivity) can be defined:

$$Accuracy = \frac{TP + TN}{TP + FP + TN + FN}$$

$$Precision = \frac{TP}{TP + FP}$$

$$Recall = \frac{TP}{TP + FN}$$

## 3.5 Optimizer and optimization tricks

During the training I used Adam, Nadam and SGD optimizers, but finally I prefered to use Adam and it helped obtained a not bad result(see Table 3).

For CXR images and COVID-19 CXR Dataset, in addition to the data preprocessing and data augmentation methods mentioned in Section 3.2, other optimization methods such as Warmup and label smoothing are also used.

Warmup is a learning rate warm-up method mentioned by He Kaiming et al[16]. Since the weights of the model in the initial epochs of training are randomly initialized, if using a large learning rate, the model may be unstable. Therefore, I used the Warmup method, using a small learning rate in the first few epochs or steps of the training, and then using the preset learning rate for training after the model is relatively stable, which makes the model convergence speed faster and more effective.

In the later stage of training, the loss of the model is reduced to a low level. At this time, if the original larger learning rate is still used, it may cause "oscillation". In order to reduce the model gradient to the local minimum as soon as possible, the learning rate should be gradually attenuated(Decay).

In combination with Warmup and Decay, I used a three-stage learning rate when training the model, that is, a smaller learning rate is used in the initial stage, and a larger preset learning rate is used in the intermediate stage after the model is stabilized, and at the later stage of training, in order to help the model further converge, a smaller learning rate is used.

In view of the obvious overfitting of the model during the training process, I also used label smoothing during training. Label smoothing was first proposed by Szegedz et al[17]. It is a regularization strategy. In Section 3.4 this paper ever introduced the cross entropy loss function for every sample $i$ with one-hot encoding label:

$$Loss_i = -\log(p_{i,j}) \quad ①$$

after using label smoothing, the one-hot encoding label will add an error term $\varepsilon$, so the label change to

$$label = \begin{cases} 1 - \varepsilon & if\ class == y \\ \frac{\varepsilon}{N-1} & if\ class\ != y \end{cases} \quad ②$$

the loss function for every sample $i$ change to

$$Loss_i = -\log[(1-\varepsilon)\,p_{i,j}] - \sum_{k \neq j} \log\left(\frac{\varepsilon}{N-1}\,p_{i,k}\right) \quad ③$$

where $N$ is the total number of classification categories, $\varepsilon$ is a small value hyperparameter, and $p_{i,j}$ is the probability that the sample $i$ is predicted to be category $j$.

Comparing ① and ③, it can be found that after using label smoothing, the loss of a single sample $Loss_i$ increases, so the global Loss will also increase accordingly, That is, a little penalty is imposed on the model classification, so the model is more robust. If the model want to achieve the loss level as before using label smooth, it must be closer to the global optimal/local optimal.

# 4 Experimental Results

## 4.1 Experiment device & environment

The configuration of experiment device is as follows:
- Graphic Card: NVIDIA GTX 1050 2GB GDDR5
- CPU: Intel(R) Core(TM) i5-7300HQ 4 cores based frequency 2.5GHz
- DRAM: Samsung 8GB DDR4 2400MHz

The enviroment and dependency of experiment is as follows:
- NVIDIA CUDA Development Kits 10.1
- Python 3.6 or higher version
- TensorFlow 2.2.0 or higher version

## 4.2 Quantitative Analysis

After training for ten epochs, the evaluation metrics of the training set, test set, and COVID-19 images in the test set are shown in Table 3.

| Metrics | Cross Entropy Loss | Accuracy | Precision | Recall |
|---|---|---|---|---|
| Train | 0.986 | 0.9325 | 0.9463 | 0.9133 |
| Test | 0.5632 | 0.873 | 0.8967 | 0.8446 |
| COVID-19 (From Test) | 0.4304 | 0.98 | 0.9798 | 0.97 |

Table 3  Metrics

It can be seen from Table 3 that the test set has an accuracy rate of 87.3%, an accuracy rate of 89.67%, and a recall rate of 84.46%. However, of the 100 COVID-19 images in the test set, 98 were correctly classified, and the recall rate and accuracy rate reached 97%. As far as evaluation metrics are concerned, the model still has space for improvement.

To show whether the model is valid or not, I write the mean and the confidence interval under 95% confidence of the predicted probability of the three types of samples in the test set in Table 4.

| Confidence | Mean | Lower Limit | Upper Limit |
|---|---|---|---|
| COVID-19 | 0.8445 | 0.8185 | 0.8705 |
| Normal | 0.6763 | 0.6462 | 0.7063 |
| Pneumonia | 0.7761 | 0.7541 | 0.7981 |

Table 4  Category prediction probability of Test Set under 95% confidence interval

It can be seen from Table 4 that under the condition of 95% confidence interval, the model has a prediction probability of more than 81% for COVID-19 samples. Based on this, it can be considered that the model is effective in identifying COVID-19, and the model has a high degree of confidence in the classification of COVID-19.

Comparing to COVID-19 samples, the model has lower confidence in classfication of the Normal and Pneumonia samples. Table 5 is the confusion matrix of the test set and Figure 4 is the visualization of the confusion matrix, which can partially explain the reason for this situation.

|  | COVID-19 | Normal | Pneumonia | Sum |
|---|---|---|---|---|
| COVID-19 | 98 | 0 | 2 | 100 |
| Normal | 9 | 197 | 36 | 242 |
| Pneumonia | 5 | 42 | 351 | 398 |
| Sum | 112 | 239 | 389 | 740 |

Tabel 5  Confuse Matrix of The Test Set

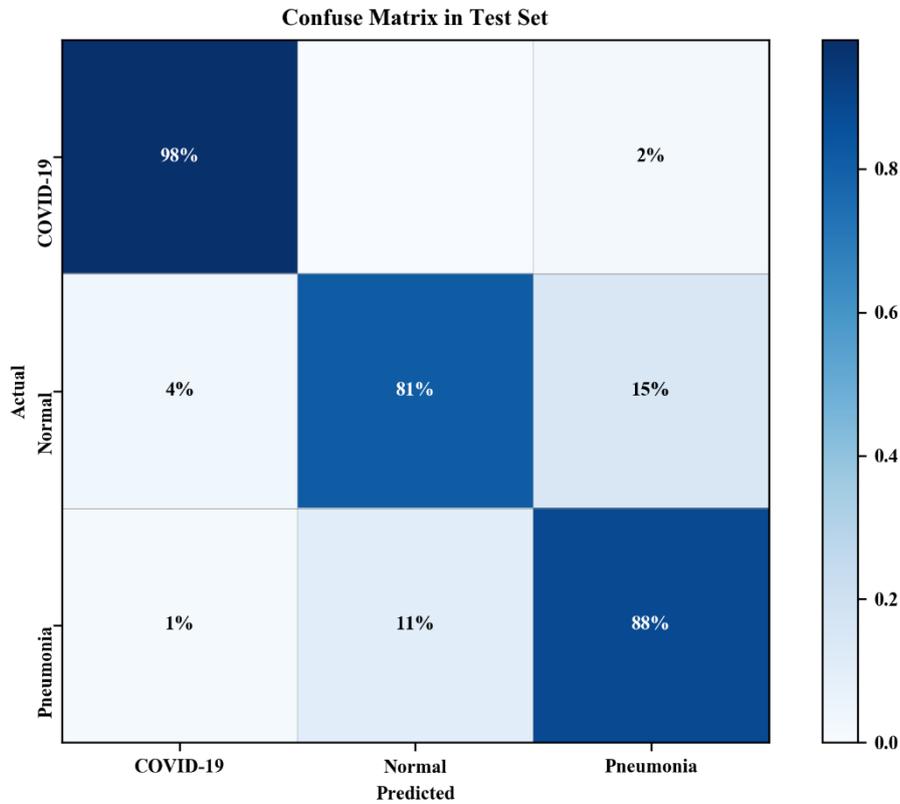

Figure4  Visualization of The Confusion Matrix in Table 5

From Table 5 and Figure 4, it can be seen that the model has a lot of confusion for the Normal samples and Pneumonia samples, the percentage of misjudgments with each other exceeded 11%. One of the possible reasons is that the class weight setting is too biased towards COVID-19. Using more reasonable class weights may further improve the performance of the model.

In addition, model has far less missed detection samples than the number of false positives samples on the COVID-19 category. Considering in real life, the cost of a false positive for COVID-19 is much less than the cost of a missed detection, therefore the model works well in COVID-19 classification.

## 4.3 Qualitative Analysis

With the help of the TensorBoard it is possible to observe trends in the evaluation metrics throughout the training process, as shown in Figure 5 and Figure 6.

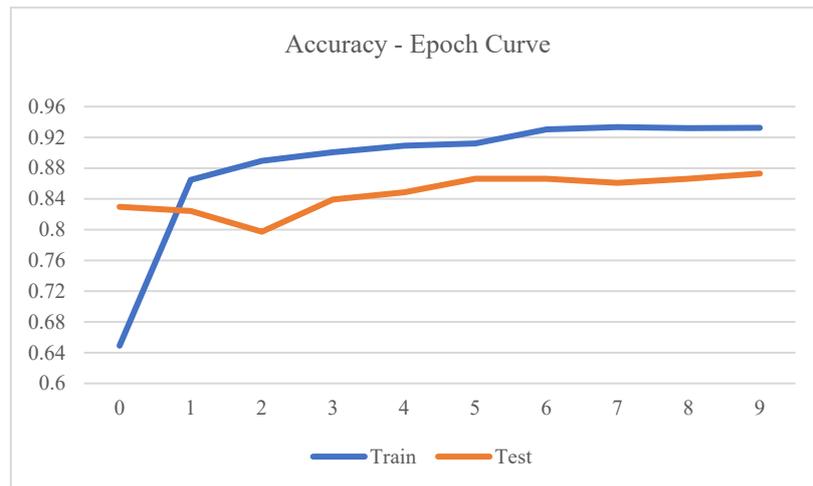

Figure 5  Accuracy-Epoch Curve

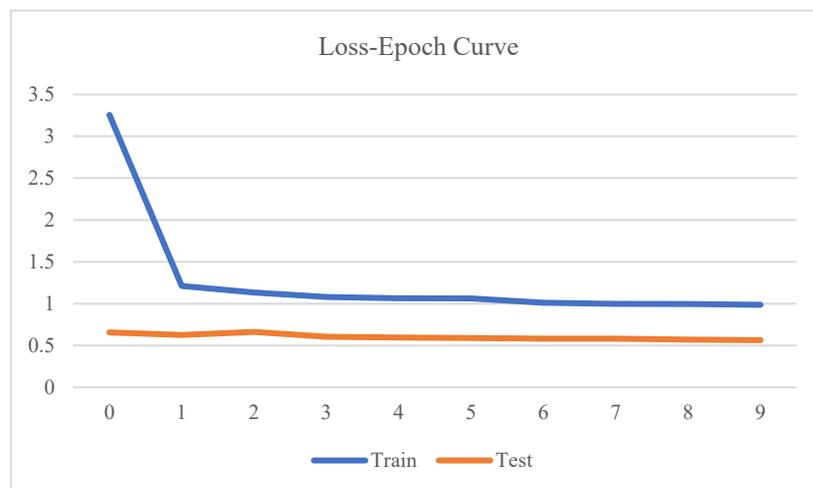

Figure 6  Loss-Epoch Curve

Observing the curve, it can be seen that as the training progresses, the loss function gradually converges, and the accuracy of the model shows an upward trend. However, the accuracy of the model on the training set and the test set has a certain distance, which shows that the model still has overfitting. So the model has a lot of space for improvement.

# 5  Conclusion

The COVID-19 epidemic situation has not been completely contained, and the adverse effects on people's lives and transportation are still continuing. During the research process, I was restricted by the research conditions, so I can only use a small image size as 80×80 to training the model, and the model has a lot of space for improvement. This paper proposed a open source COVID-19 dataset named COVID-19 CXR Dataset, and proposed a DCNN model achieved 87.3% accuracy, 89.67 % precision, and 84.46% recall, and correctly classifies 98 out of 100 COVID-19 x-ray images in the test set. This paper can be used by other researchers to conduct more in-depth research based on this paper. I hope to work with other researchers to make small contributions to contain the epidemic and industry development. Finally, I hope the epidemic will end soon and return to normal life!

# Reference

[1] World Health Organization. Coronavirus disease 2019 (COVID-19): situation reports, 161[J/OL]. 2020. https://www.who.int/docs/default-source/coronaviruse/situation-reports/20200629-covid-19-sitrep-161.pdf?sfvrsn=74fde64e_2

[2] National Health Comission of the People's Republic of China. Diagnosis and treatment of pneumonia with new coronavirus infection (trial version 7)［EB/OL］. [2020-03-04]. http://www.nhc.gov.cn/yzygj/s7653p/202003/46c9294a7dfe4cef80dc7f5912eb1989.shtml

[3] National Health Comission of the People's Republic of China. Diagnosis and treatment of pneumonia with new coronavirus infection (trial version 5)［EB/OL］.［2020-02-15］. http://www.nhc.gov.cn/yzygj/s7653p/202002/3b09b894ac9b4204a79db5b8912d4440.shtml

[4] DENG Liangna, et al. Application progress of imaging in COVID-2019 [J/OL]. Chinese Journal of Medical Physics. 2020, http://kns.cnki.net/kcms/detail/44.1351.r.20200511.0905.002.html

[5] SHI Heshui, et al. Radiologic Features of Patients with 2019-nCoV Infection[J]. Journal of Clinical Radiology. 2020, DOI:10.13437/j.cnki.jcr.20200206.002

[6] Annarumma M, Withey S J, Bakewell R J, et al. Automated triaging of adult chest radiographs with deep artificial neural networks[J]. Radiology, 2019, 291(1): 196-202.

[7] Pengpai News, Alibaba Cloud AI diagnosis new technology: new coronary pneumonia CT image recognition accuracy rate is 96% [EB/OL]. [2020-02-15]. https://www.thepaper.cn/newsDetail_forward_6000556

[8] Xu X, Jiang X, Ma C, et al. Deep Learning System to Screen Coronavirus Disease 2019 Pneumonia[J]. arXiv preprint arXiv:2002.09334, 2020.

[9] Hemdan E E D, Shouman M A, Karar M E. Covidx-net: A framework of deep learning classifiers to diagnose covid-19 in x-ray images[J]. arXiv preprint arXiv:2003.11055, 2020.

[10] Wang L, Wong A. COVID-Net: A Tailored Deep Convolutional Neural Network Desi

gn for Detection of COVID-19 Cases from Chest X-Ray Images[J]. arXiv preprint arXiv:2003.09871, 2020.

[11] Cohen J P, Morrison P, Dao L, et al. COVID-19 Image Data Collection: Prospective Predictions Are the Future[J]. arXiv preprint arXiv:2006.11988, 2020.

[12] Chung, A. Figure 1 COVID-19 chest x-ray data initiative[EB/OL]. https://github.com/agchung/Figure1-COVID-chestxray-dataset.

[13] Chung, A. Actualmed COVID-19 chest x-ray data initiative[EB/OL]. https://github.com/agchung/Actualmed-COVID-chestxray-dataset.

[14] P. Mooney. Chest X-Ray Images (Pneumonia)[EB/OL]. https://www.kaggle.com/paultimothymooney/chest-xray-pneumonia

[15] Simonyan K, Zisserman A. Very deep convolutional networks for large-scale image recognition[J]. arXiv preprint arXiv:1409.1556, 2014.

[16] He K, Zhang X, Ren S, et al. Deep residual learning for image recognition[C]//Proceedings of the IEEE conference on computer vision and pattern recognition. 2016: 770-778.

[17] Szegedy C, Vanhoucke V, Ioffe S, et al. Rethinking the inception architecture for computer vision[C]//Proceedings of the IEEE conference on computer vision and pattern recognition. 2016: 2818-2826.